\documentclass[12pt]{article}
 \usepackage{amssymb}
\usepackage{amsfonts}
 \usepackage{amsmath}
 \usepackage[mathscr]{eucal}
 \usepackage{amsthm}
 \usepackage{bbold}
 \usepackage{bm}
 \usepackage{graphicx}
 \usepackage[T2A]{fontenc}

\newcommand{\be}{\begin{equation}}
\newcommand{\ee}{\end{equation}}
\newcommand{\bea}{\begin{eqnarray}}
\newcommand{\eea}{\end{eqnarray}}

\begin{document}

\begin{titlepage}

\vspace*{3cm}

\begin{center}
{\LARGE\bf Covariant quantization }

\vspace{0.5cm}

{\LARGE\bf  of $d = 4$ Brink-Schwarz superparticle }

\vspace{0.5cm}

{\LARGE\bf with Lorentz harmonics}

\vspace{2cm}

{\large\bf V.G. Zima,\,\, S.A. Fedoruk}

\vskip 1cm

\ {\it Kharkov State University, Kharkov, Ukraine} \\
{\tt fedoruk@theor.jinr.ru}

\end{center}

\vspace{2cm}

\begin{abstract}
\noindent Covariant first and second quantization of the free $d=4$ massless superparticle are implemented with the introduction
of purely gauge auxiliary spinor Lorentz harmonics.
It is shown that the general solution of the condition of maslessness is a sum of two independent chiral superfields
with each of them corresponding to finite superspin.
A translationally covariant, in general bijective correspondence between harmonic and massless superfields is constructed.
By calculation of the commutation function it is shown that in the considered approach only harmonic fields with correct connection between
spin and statistics and with integer negative homogeneity index satisfy the microcausality condition.
It is emphasized that harmonic fields that arise are reducible at integer points.
The index spinor technique is used to describe infinite-component fields of finite spin;
the equations of motion of such fields are obtained, and for them
Weinberg's theorem on the connection between massless helicity particles and the type of nongauge field
that describes them is generalized.
\end{abstract}

\vspace{0.8cm}

\newpage

\end{titlepage}
\setcounter{footnote}{0}
\setcounter{equation}{0}

\section*{Introduction}

In this paper, we implement the covariant quantization of a massless superparticle [1] in ordinary four-dimensional
space-time in the framework of the harmonic approach. Since a superparticle can be regarded as the point limit of a superstring,
the theory of which may give a consistent relativistic quantum theory, the problem of the covariant quantization of a
superparticle has fundamental importance and has been investigated in a large number of studies. Common features of these
studies are the use of auxiliary variables and -- postulated or obtained in some manner -- the Cartan--Penrose representation
for the isotropic energy--momentum vector of the panicle in a form proportional to a product of spinors. At the classical level,
the differences relate to the method used to introduce gauge symmetries, which are needed to eliminate the auxiliary variables
on the transition to the physical gauge.

In the twistor approach, one seeks variables in which the theory admits a formulation in which second-class constraints
are absent and there are sufficient first-class constraints to gauge away the unphysical degrees of freedom. The commuting
spinors in terms of which the energy--momentum vector is expressed are dimensional. Depending on the variables that in
conjunction with these spinors form the supertwistor, the formulations with the Ferber [2,3] or Berkovits [4,5] supertwistors
are obtained.

In the harmonic approach, the postulated group structure of the necessary gauge symmetries is in the foreground. The
corresponding auxiliary variables (harmonics) are dimensionless and parametrize a homogeneous space, which, as a rule, is a
factor space of the Lorentz group with respect to some subgroup. Depending on the choice of this subgroup, if it is not dictated
by the problem, one can distinguish compact and noncompact harmonics [6--9].

Of course, there can also be a hybrid twistor--harmonic approach when a transparent harmonic group structure is a
consequence of a felicitously chosen action, which is written initially in terms of harmonics with the twistor representation for
the energy--momentum vector mentioned above [10, 11].

The first success in the problem of covariant quantization of a superparticle in ten-dimensional space-time was achieved
in [12] using vector Lorentz harmonics [6], two of which were expressed in terms of Lorentz spinors. The need to introduce
these last was also emphasized in [13].

When the classically equivalent actions for the massless superparticle are quantized, there arise ordering constants, the
values of which determine the helicity and the method of description of the superparticle. At the same time, it is well known
that the infinite-component fields that correspond to almost all values of the ordering constants need not necessarily satisfy the
requirement of locality (microcausality) and the theorem on the connection between the spin and the statistics. This paper is
devoted to a discussion of these questions in the framework of a consistent covariant quantization of the harmonic superparticle.

The harmonic variables are introduced by adding to the standard Lagrangian of the Brink--Schwarz superparticle [1] both
a kinetic term for these variables and a combination of constraints that generate gauge transformations permitting the harmonics
to be regarded as purely gauge auxiliary variables. The introduced harmonics play the role of a bridge between the ordinary
spinor basis and the light-cone basis. A canonical transformation to isotropic coordinates makes it possible, after adaptation
of the isotropic basis to the 4-momentum of the superparticle (a certain gauge fixing), to separate the first- and second-class
constraints covariantly. This procedure represents a covariantization of the light-cone gauge by means of the auxiliary variables.

The first operator quantization in the light-cone basis is done in the coordinate representation, in which chiral variables
are introduced to ensure commutativity of the coordinates. The condition of masslessness, imposed as a first-class constraint
on the wave function, has a general solution in the form of a sum of two independent terms, which depend on the fight and left
chiral coordinates, respectively. In application to one of them, the remaining harmonic first-class constraints generate the
minimal parabolic subgroup of the Lorentz gauge group, in accordance with which the quantum-mechanical space of the
harmonics is the compact homogeneous space $\mathbf{CP}^1$. Therefore, the compactness of the harmonics is not the result of an arbitrary
choice but follows from the symmetry of the problem.

Restricting ourselves to the consideration of a superfield of definite chirality, and fixing certain gauges, we obtain as a
result of the first operator quantization the harmonic superfield $\Psi^{c,s}(x_L^{++}, \theta^+, v^+)$, which depends on the left isotropic
supercoordinates $x_L^{++}$, $\theta^+$ and the complex harmonic spinor $v^+$, where the harmonic superscript + determines the
transformation law with respect to the gauge group $SO{\uparrow}(1,1)\oplus SO(2)$. The constraints imposed on the harmonic superfield that
generate this subgroup contain ordering constants -- the conformal weight $c$ and the spin weight $s$. By virtue of the requirement
that the field be single valued, the latter must be an integer or half-integer (see, for example, [14]).

Calculation of the supersymmetric generalization of the Pauli--Luba\'{n}ski pseudovector shows that the superhelicity is the
opposite of the spin weight of the harmonic superfield and the component fields describe massless particles of helicities $-s$ and
$-s+ 1/2$. Expressing the isotropic variables in the obtained superfield in terms of the ordinary superspace coordinates, we
obtain a field in which the role of the index spinor is played by the harmonic spinor $v^+$ and which satisfies harmonicity
conditions ensuring the above-mentioned dependence on the isotropic variables. An infinite-dimensional representation of the
Lorentz group is realized on the index of this superfield.

The finiteness of the spin in theories with twistor representation of the 4-momentum is well known [15--17].

To clarify the physical consequences of the microcausality condition, we first consider massless fields on whose index an
arbitrary, not necessarily finite-dimensional, irreducible representation of the Lorentz group is realized with particles of finite
spin. This required the use of an adequate technique, in particular an index-free expression of multicomponent and infinite-
component fields as homogeneous functions of the index spinor $\zeta$. The second quantization of the massless fields is done in
the framework of the usual formalism. The Wigner wave function obtained in the momentum representation shows that the spin
weight of the nongauge field that describes massless particles of finite spin is necessarily equal to their helicity. For finite-
component fields, this result was obtained in [18]. We find equations of motion without a restriction on the irreducible
representation of the Lorentz group realized on the index. These equations are in a certain duality with the harmonicity
condition.

Calculation of the commutation function for the massless fields shows that the condition of locality (microcausality) can
be achieved only if there is a correct connection between the spin and the statistics and only for fields on the index of which
a finite-dimensional representation of the Lorentz group is realized (integer positive values of the homogeneity index [19]).

Substitution in the second-quantized field of the momentum expressed in the light-cone basis adapted to it represents it as
the result of an integral transformation of some harmonic superfield with spin and conformal weights corresponding to the
opposite homogeneity index. For scalar fields, such a procedure is described in [9]. The kernel of the transformation is a
homogeneous function of the contraction of the index spinor $\zeta$ with the harmonic $v^+$. This makes it possible to formulate a
prescription for second quantization of the harmonic superfields. On the other hand, there is a unique method of bijectively
and translationally invariantly associating a harmonic field with an ordinary field. This uses the integral representation described
in [9] and establishes the equivalence of representations of the Lorentz group in spaces of homogeneous functions.

In the most interesting case of integer homogeneity index, when the massless fields associated with the harmonic
superfields can have finitely many components, the first quantization gives harmonic fields that transform in accordance with
irreducible representations. In the description of elementary particles, this leads to the necessity of either restricting
consideration to a wave function in an invariant subspace or introducing an additional ``gauge'' invariance that gives an arbitrary
correction to the field in the form of a function from such an invariant subspace. However, the invariant subspace
corresponding to positive homogeneity index of the harmonic field is finite dimensional and is therefore not suitable for
describing physical massless fields. In the remaining cases, only fields with positive integer index satisfy the condition of
microcausality among the massless fields that are associated in the manner described above with the harmonic field, in the case
of noninteger homogeneity indices, and the gauge classes and fields from the invariant subspaces for integer indices. To such
fields with positive integer index there correspond gauge classes of harmonic fields with negative integer homogeneity index.

Of course, the fact that the fields with noninteger values of the homogeneity index do not satisfy the microcausality
condition does not mean that such fields cannot arise from the quantization of extended objects (strings, membranes). Like the
difficulties in the covariant quantization of a superparticle, this could be a consequence of the fact that in a consistent quantum
theory the fundamental objects are extended and they are not the point limits of such objects, i.e., particles. In the occurrence
of infinite-component nonlocal fields in the first quantization of the superparticle one can see a manifestation of the circumstance
that the theory of point objects is the limiting case of a consistent relativistic quantum theory of extended objects, say superstring
theory. As a result of this, a fairly developed and consistent theory of the superparticle must necessarily affect some aspects
of superstring theory.

In our opinion, the formal arguments advanced in [7--9] are not sufficient to ``prohibit'' noncompact Lorentz harmonics,
which are possible, because of the noncompactness of the Lorentz group, together with compact harmonics (see also [4]). We
are justified in using, with the necessary care, auxiliary variables that can parametrize both compact and noncompact spaces.
However, the claim of implementation of covariant quantization in studies using compact harmonics is not sufficiently justified
before these studies are concluded with the construction of a covariant S matrix with specification of the appropriate asymptotic
conditions. In this sense, the results obtained in this way must be regarded as preliminary. In the framework of the requirement
that the wave function be single valued, compactness of the harmonics frees the theory of this shortcoming.

The considered approach to superparticle quantization by the introduction of purely gauge auxiliary variables, which is
natural from the point of view of the philosophy of BFV--BRST quantization [20--22], makes it possible to exploit fully the
transparent group structure of the Lorentz harmonics and control the employed gauge symmetries and the start of the
quantization procedure. The approaches that begin with an arbitrary twistor-harmonic action, which in our formalism represents
the result of a certain choice of gauge, are forced to recover the constraints by a necessary preliminary investigation.

As a first step to the construction of a quantum theory of interactions, one should consider the description of a superparticle
in a background Yang--Mills and/or gravitational field. This requires the development of a procedure of ``soft'' -- not leading
to physical consequences -- introduction at the classical level of auxiliary harmonic coordinates for the background fields. Then
the addition of the auxiliary coordinates will be done consistently for all physical objects of the theory. Naturally, this procedure
assumes an adequate change in the composition of the superfields of the background and of the constraint conditions that they
satisfy, and also the specification of a gauge that makes it possible to dispense with the auxiliary coordinates. Of course, since
the harmonics are a classical basis for the description of spin, the adjoining of them introduces, despite their purely gauge
nature, an irreversible change into the system that is clearly manifested as spin after quantization -- a change in the topology
of the phase space. As a result of the introduction of the harmonics, the fields themselves must acquire additional excitations
propagating in the added directions. The corresponding masses are large (compactness of the harmonic space), but this does
not prevent their contribution from competing under certain conditions with the graviton contribution [23]. It is clear that these
reservations are not important for the consideration of a particle in a background, i.e., in a fixed configuration of classical fields.
The stipulations merely delineate the immediate vicinity of the theory of the superparticle in the more general theory that
contains it: The construction of the theory of the superparticle is a model problem of superstring theory.

The introduction of the harmonics makes it possible for the particle to ``feel'' not only the frame coefficient but also the
Lorentz connection. More precisely, in the harmonic superspace in the spirit of Kaluza--Klein theories [23] there is a place
among the frame components for not only the ``original'' ones and those introduced for the harmonic sector but also for the
``original'' Lorentz connection. It is in this way that the interaction of internal (not spin) degrees of freedom with a background
Yang--Mills field is usually considered [24].

Let us consider now the interaction with the gravitational background. The frame coefficients that arise in the considered
manner, $E^{\mathcal{M}}_{\alpha}$ and $E^{\mathcal{M}}_{\dot\alpha}$, 
where $\mathcal{M}$ labels all coordinates of the harmonic superspace, contain all the fields of $N= 1$ simple
supergravity even after they have been made to satisfy (suitably generalized) nonconventional constraints of the second type in
the classification of [251. It remains to express, by imposing conventional constraints (of the first type [25]), the remaining
frame coefficients and connections in terms of the chosen fundamental ones, confirming that the flat superspace structure is
reproduced in the tangent space of the Lorentz basis and eliminating by the choice of an appropriate constraint of the third type
the compensator of the Weyl transformations as one would an independent superfield. In the harmonic sector, the frame form
is determined by the structure of the group, while the connection is determined by the ``arbitrary'' choice of the torsion
(curvature), which completes the set of conventional constraints. By the use of the Bianchi identities (see, for example, [26]),
the system of constraints obtained in this way can be tested for independence, and by resolving the constraints [27] one can
obtain a formulation like that of [28], in which the real physical superspace is a hypersurface in a complex superspace and it
is precisely the harmonic coordinates obtained on quantization of the harmonic superparticle in flat space that play the role of
Gaussian coordinates on the surface. A detailed discussion of the mentioned problems is the subject of a separate paper.

There is an extensive literature on (super)particles, both free and in background fields. While making no claim to giving
a complete list of these studies, we mention publications in this journal that are closest to the considered questions: particle in
a background [29], particle with spin [30], superparticle [31], two-point functions [32], particle with torsion [33].

In the paper, we use the standard notation of [34].

\section{Classical treatment}

We write the covariant form in superspace with coordinate $x^{\mu}$, $\theta^{\alpha}$, $\bar {\theta}^{\dot {\alpha}}$ in the form
$$
\omega_\theta^{\mu}=dx^{\mu}-i\theta\sigma^{\mu}\bar
{\theta}+i\theta\sigma^{\mu}\,d\bar {\theta}\,.
$$
On the worldline of the superparticle, $\omega=\dot{\omega}\,d\tau$, where $\tau$ is an evolution parameter.
In the first-order formalism, the Lagrangian in the action
$
A=\int Ld\tau
$
of the massless superparticle has the form \cite{1}
\begin{equation}
L=p\dot\omega_\theta-\frac e2 \,p^2\,,
\end{equation}
where $e$ is a Lagrange multiplier.  In the Hamiltonian formulation, the momenta and coordinates are subject to the constraints
\begin{equation}
p^2\approx 0\,,
\end{equation}
\begin{equation}
d_{\theta}\equiv i p_{\theta}+\hat {p}\overline {\theta}\approx 0\,,
\qquad
\bar d_{\theta}\equiv i \bar p_{\theta}+\theta\hat {p} \approx 0\,,
\end{equation}
between which the nonvanishing Poisson brackets are
\begin{equation}
\{d_{\theta},\bar {d_{\theta}}\}_{PB}=2i\hat p\,,
\end{equation}
where $\hat {p}\equiv p^{\mu}\sigma_{\mu}$.

When the third axis is directed along the three-dimensional momentum of the superparticle, $\hat {p}=\mathrm{diag}(2p^0,0)$. Therefore,
half of the Fermi constraints (3) are, like the condition of masslessness, first class, and half are second class. A covariant
separation of the Fermi constraints (3) into classes cannot be done without the use of auxiliary variables, and this hinders the
implementation of a covariant quantization procedure.

Among the various sets of auxiliary variables, Lorentz harmonics [6--13] are particularly attractive on account of the
simplicity with which their gauge nature can be ensured.

In the case d=4 considered here, the Lorentz harmonics are Bose spinors
$$
v^\pm_{\alpha}\,, \qquad\bar v^\pm_{\dot\alpha}=\overline{v^\pm_{\alpha}}\,.
$$
The spinors $v^+$ and $v^-$, and also $\bar v^+$ and $\bar v^-$, can be regarded as the columns of two complex mutually conjugate $2\times 2$ matrices $v$ and $\bar v$. The kinematic constraints
\begin{equation}
h\equiv v^+v^-+1\approx 0\,,\qquad \bar h\equiv \bar v^-\bar v^+ +1\approx 0\,,
\end{equation}
which make the determinants of these matrices have the value unity, transform them into elements of the group $SL(2,\mathbb{C})$ -- the
universal covering of the proper orthochronous Lorentz group $SO{\uparrow}(3,1)$. We can therefore write in the standard manner a
system of first-class constraints that form the algebra of the group $SL(2,\mathbb{C})$ and preserve the kinematic constraints (5):
\begin{equation}
d^{\pm\pm}=v^\pm p^\pm_v\approx 0\,,\qquad \bar d^{\pm\pm}\equiv \bar v^\pm \bar p^\pm_v\approx 0\,,
\end{equation}
\begin{equation}
d^0=v^+p^-_v-v^-p^+_v -k.c.\,,\qquad \overline{d^0}=-d^0\,,
\end{equation}
\begin{equation}
d^{+-}=v^+p^-_v-v^-p^+_v + k.c.\,,\qquad \overline{d^{+-}}=d^{+-}\,.
\end{equation}
Here, $p^\pm_v$ and $\bar p^\pm_v$ are the momenta canonically conjugate to the harmonics $v^\mp$ and $\bar v^\mp$.
Naturally, the number of constraints
in the system (6)--(8) is equal to the dimension of the Lorentz group in $d=4$.

The constraints $d^{+-}$ and $d^0$ generate local $SO{\uparrow}(1,1)$ and $SO(2)$) transformations, respectively. Obviously
$$
\{v^\pm,d^0\}_{PB}=\{v^\pm,d^{+-}\}_{PB}= \pm v^\pm\,.
$$
The harmonic indices $\pm$ are collective: They indicate the signs of the conformal and spin weights, which are equal to the
harmonic index of the variables $v^\pm$, $p^\pm_v$; the sign of the spin weight of the conjugate quantities $\bar v^\pm$, $\bar p^\pm_v$
the opposite of the
harmonic index. The weights of quantities with several harmonic indices are found by adding the weights corresponding to the
individual indices. In determining the sign of a spin weight corresponding to a harmonic index it is necessary to take into
account the kind of index -- dotted or undotted -- with which it is associated.

The gauge conditions
\begin{equation}
\chi\equiv v^-p^+_v+v^+p^-_v\,,\qquad
\bar \chi\equiv \bar v^-\bar p^+_v+\bar v^+\bar p^-_v
\end{equation}
are conserved by the harmonic constraints (6)--(8) and are admissible with respect to the kinematic constraints (5):
$\{h,\chi\}_{PB}=2(h-1)$ and c.c. In what follows, the kinematic constraints will be assumed to be satisfied in the strong sense. The Dirac
brackets [35] of the ``covariant momenta'' -- the harmonic constraints d and the harmonics $v$ and $\bar v$ themselves -- are identical
to the Poisson brackets, since these quantities commute with the kinematic constraints (5). The Dirac brackets with the
participation of harmonic momenta will be considered when necessary. We mention in passing that for the harmonic momenta
and harmonics of different chiralities the Dirac brackets are identical to the Poisson brackets.

The spinor basis of the Lorentz harmonics, for which the completeness condition has the form
$$
v^-_\alpha v^+_\beta-v^+_\alpha v^-_\beta=\epsilon_{\alpha\beta}
$$
and c.c., determines a complex isotropic tetrad
\begin{equation}
\begin{array}{c}
u_{\mu}^{\pm\pm}=v^{\pm}\sigma_{\mu}\bar {v}^{\pm}\,,\qquad
\bar u_{\mu}^{\pm\pm}= u_{\mu}^{\pm\pm}\,, \\ [6pt]
u_{\mu}^{\pm\mp}=v^{\pm}\sigma_{\mu}\bar {v}^{\mp}\,,\qquad
\bar u_{\mu}^{\pm\mp}= u_{\mu}^{\mp\pm}\,.
\end{array}
\end{equation}
The nonvanishing scalar products of the tetrad vectors are
$$
u^{--}u^{++}=-2\,, \qquad u^{+-}u^{-+}=+2\,,
$$
and the completeness condition (decomposition of the identity) is
$$
-u^{++}_{(\mu} u^{--}_{\nu)}+u^{+-}_{(\mu} u^{-+}_{\nu)}=\eta_{\mu\nu}\,,
$$
where the round brackets denote symmetrization with respect to the enclosed indices. The vectors of the tetrad constructed in
this manner, regarded as rows, form a complex unimodular matrix that connects the orthonormal frame to the isotropic frame
while preserving the direction of the time axis ($u^{++}_0>0$).

In the action
$$
A_{aux}=\int L_{aux}d\tau
$$
 for the harmonic phase variables, the Hamiltonian is a linear combination of the harmonic
constraints, the coefficients of which are Lagrangian multipliers. By construction, the harmonic superparticle with action
\begin{equation}
A_{harm}=A+A_{aux}
\end{equation}
is classically equivalent to the original superparticle (1). However, for it one can, due to the presence of the auxiliary harmonic
variables, consider without loss of covariance the coordinates
\begin{equation}
x^{\pm\pm}=x^{\mu}u_{\mu}^{\pm\pm}\,, \qquad x^{\pm\mp}=x^{\mu}u_{\mu}^{\pm\mp}\,,
\qquad \theta^\pm=\theta v^\pm\,, \qquad\bar{\theta}^\pm=\bar v^\pm\bar\theta
\end{equation}
with respect to the complex isotropic tetrad adopted to the 4-momentum. We shall call these coordinates isotropic, or
coordinates in the light-cone basis (in the light-cone gauge). The transition to the light-cone variables is a canonical
transformation. For the superspace momenta of the isotropic coordinates, we have the analogous expressions
$$
p^{\pm\pm}=p^{\mu}u_{\mu}^{\pm\pm}\,, \qquad p^{\pm\mp}=p^{\mu}u_{\mu}^{\pm\mp}\,,
\qquad p_\theta^\pm=\pm v^\pm p_\theta \,, \qquad\bar p_{\theta}^\pm=\mp\bar v^\pm\bar p_\theta \,;
$$
the harmonic momenta transform inhomogeneously. In terms of the isotropic variables, the superparticle constrains (2)--(3)
take the form
\begin{equation}
p^{2}=-p^{--}p^{++}+p^{+-}p^{-+}\approx 0\,,
\end{equation}
\begin{equation}
\begin{array}{l}
d^\pm=v^\pm d=\mp i p_\theta^\pm-\bar\theta^- p^{\pm +}+\bar\theta^+ p^{\pm -}\approx 0\,, \\ [6pt]
\bar d^\pm= d \bar v^\pm =\mp i \bar p_\theta^\pm- \theta^- p^{+\pm}+\theta^+ p^{-\pm}\approx 0\,.
\end{array}
\end{equation}
In the light-cone basis, the harmonic constraints (6)--(8) acquire additional terms that ensure their action on the light-cone
variables:
\begin{equation}
\begin{array}{l}
d^{\pm\pm}_c=d^{\pm\pm}+\frac12
\left( x^{\pm\pm}p^{\pm\mp}-x^{\pm\mp}p^{\pm\pm}\right)
 -\theta^\pm p_\theta^{\pm}\approx 0\,,\qquad  \bar d^{\pm\pm}_c=\overline{d^{\pm\pm}_c}\,,\\ [6pt]
d^{+-}_c=d^{+-}+\sum_\pm (\mp) \left( x^{\pm\pm}p^{\mp\mp}
+\theta^\pm p_\theta^{\mp}+\bar\theta^\pm \bar p_\theta^{\mp}\right)\approx 0\,, \\ [6pt]
d^{0}_c=d^{0}+\sum_\pm (\pm) \left( x^{\pm\mp}p^{\mp\pm}
-\theta^\pm p_\theta^{\mp}+\bar\theta^\pm \bar p_\theta^{\mp}\right)\approx 0\,.
\end{array}
\end{equation}
Of course, the harmonic constraints (15) and the constraint (13) are, as before, first class, while the Fermi constraints (14) are
a mixture of first- and second-class constraints:
$$
\{d^\pm, \bar d^\pm \}_{PB}=2i p^{\pm\pm}\,,\qquad  \{d^\pm, \bar d^\mp \}_{PB}=2i p^{\pm\mp}\,.
$$

Without loss of covariance, we can now adapt the light-cone basis to the superparticle momentum, imposing the gauge
conditions 
$$
p^{+-}\approx 0\,,\qquad p^{-+}\approx 0\,.
$$
Such a gauge is admissible for the boost harmonic constraints $d_c^{\pm\pm}$ and $\bar{d_c}^{\pm\pm}$, since
$$
\{p^{\pm\mp}, d_c^{\pm\pm} \}_{PB}=\{p^{\pm\mp}, \bar{d_c}^{\pm\pm} \}_{PB}= p^{\pm\pm}\,.
$$
Because of the relation $p^{2}\approx 0$, this leads to the constraint
$$
p^{\pm\pm}\approx 0
$$
for$p^{\mp\mp}\neq 0$. Thus, by the boost transformations $d_c^{\pm\pm}$ and $\bar{d_c}^{\pm\pm}$ one of the isotropic vectors of the isotropic tetrad is directed along the 4-momentum of the superparticle:
$$
p_\mu \approx -\frac12\, p^{\mp\mp}u_\mu^{\pm\pm}\,.
$$
As a result, the matrix of Poisson brackets of the Fermi constraints is ``diagonalized'', and the constraints
are covariantly separated accordingly to classes with $d^\mp$ and $\bar d^\mp$ playing the role of the second-class constraints and
$d^\pm$ and $\bar d^\pm$ the role of the first-class constraints. This procedure for separating the constraints is a covariantization by means of the
harmonics of the noncovariant analysis of the Fermi constraints (3) made at the beginning of this section.

It is known [36] that the first-class constraints can be covariantly separated from (3) by forming the combination $\phi\equiv \hat p d \approx 0$
and c.c. The system of first-class constraints obtained in this manner is redundant: $$
\bar \phi\hat p-dp^2\approx 0\,,
$$
 and it has infinite degree of
reducibility. The presence of the auxiliary variables makes it possible to avoid the redundancy by separating as first-class
constraints one covariant isotropic component from each of the constraints $\phi$ and $\bar\phi$:
$\phi^\mp \approx 0$ and c.c. Then the remaining
constraints $d^\mp$ and $\bar d^\mp$ are second class. This analysis reduces to the previous analysis in the gauge $p^{+-}\approx 0$ and $p^{-+}\approx 0$ for
$p^{\mp\mp}\neq 0$ since then the constraints
$$
\phi^\mp \equiv \mp p^{\mp\mp}\bar d^\pm \pm p^{\mp\pm}\bar d^\mp
$$
and c.c.
are equivalent to the constraints $\bar d^\pm$ and $d^\pm$.

\section{First operator quantization}

Before the first operator quantization of the harmonic superparticle (12) with the constraints (13)--(15) is performed, it
is necessary to gauge away some variables, realizing thereby a choice of the quantum gauge subgroup by matching the complex
isotropic tetrad to the momentum of the superparticle. Namely, the following gauge conditions are always admissible:
\begin{equation}
p^{+-}\approx 0\,,\qquad p^{-+}\approx 0\,.
\end{equation}
The remaining gauge subgroup is the minimal parabolic (Borel) subgroup of $SL(2,\mathbb{C})$. Fixing the gauge with respect to all boost
transformations except the transformations in $SO{\uparrow}(1,1)$ would give a subgroup smaller than parabolic and would lead to
noncompact harmonics. At this stage, explicit separation of the independent variables is not expedient.

We shall perform the quantization in the coordinate representation, in which the coordinate operators are diagonal and the
momenta are realized in terms of operators of differentiation.

Not all the coordinates $x^{++}$, $\theta^{\pm}$,  $\bar{\theta}^{\pm}$, $v^\pm$, $\bar{v}^\pm$ that remain after the gauge conditions (16) have been imposed and the
coordinates $x^{\pm\mp}$ eliminated will commute; this is due to the contribution of the second-class Fermi constraints to the Dirac
brackets. It is convenient to restore commutativity by the change of coordinates
$$
x_L^{++}\equiv x^{++}-2\imath\theta^+\bar\theta^+\,,\qquad
x_R^{--}\equiv x^{--}+2i\theta^-\bar\theta^-\,,
$$
which reduces the Fermi constraints (14) in the gauge (16) to the form
\begin{equation}
d^+|\equiv  -i p_\theta^+ \approx 0\,, \qquad
\bar d^+|\equiv   - i \bar p_\theta^+ - 2\theta^- p_R^{++}\approx 0\,,
\end{equation}
\begin{equation}
d^-|\equiv   i p_\theta^\pm - 2\bar\theta^+ p_L^{--}\approx 0\,, \qquad
\bar d^-|\equiv   i \bar p_\theta^-\approx 0\,,
\end{equation}
where the vertical bar denotes the use of the imposed gauge conditions. The chiral momenta canonically conjugate to the chiral
coordinates $x_L^{++}$ and $x_R^{--}$, they are $p_L^{--}$ and $p_R^{++}$, commute with each other and with all the considered coordinates.
Therefore, they can be realized as operators of differentiation with respect to the chiral coordinates:
\begin{equation}
p_L^{--}=2i{\partial\over{\partial x_L^{++}}}\,,\qquad
p_R^{++}=2i{\partial\over{\partial x_R^{--}}}\,.
\end{equation}
It is always clear from the context in what sense the ``hat'' is used: to denote an operator, as in (19), or a contraction of a 
4-vector with $\sigma$-matrices, as in (3).

The condition of masslessness (13), expressed in terms of the chiral variables with allowance for (16),
\begin{equation}
-p^2|\approx p_L^{--}p_R^{++}\approx 0\,,
\end{equation}
leads, when imposed as a first-class constraint on the wave function, to the equation
\begin{equation}
\frac{\partial^2 \Psi}{\partial x_L^{++}\partial x_R^{--}}= 0\,,
\end{equation}
whose general solution has the form
\begin{equation}
\Psi= \Psi_L(x_L^{++})+\Psi_R(x_R^{--})\,.
\end{equation}
With respect to the chiral isotropic coordinates, these are two plane waves propagating in opposite directions. The rays
corresponding to them are opposite generators of the light cone.

In (22), the terms can be considered independently, and therefore we shall in what follows for definiteness consider the
first of them:
$$
\Psi= \Psi_L\,.
$$
For states with $p\neq 0$, this means that
$$
p_R^{++}=0\,,
$$
while $p_L^{--}\neq 0$. Under these conditions, it is a
consequence of (20) in the classical theory that one could regard the vanishing of $p_R^{++}$ as a first-class constraint for which the
gauge condition
$$
x_R^{--}\approx 0
$$
is admissible. At the same time, the gauge conditions (16) on the transverse components of the
momentum of the superparticle are already definitely admissible for the constraints $d_c^{--}$ and $\bar d_c^{--}$,
and this allows us to regard
the coordinates $x^{+-}$ and $x^{-+}$ as not independent. It is also determined which Fermi constraints play the role of second-class
constraints -- they are $d_\theta^-$ and $\bar d_\theta^-$, and this allows us to regard $\bar\theta^+$ and $\bar p_\theta^-$
as nonindependent variables. The introduction of
the corresponding Dirac brackets does not change the commutation relations for the remaining variables.

Thus, the considered solution of the condition of masslessness is that the wave function $\Psi=\Psi_L$  is a function of the
(anti)commuting coordinates
$$
x_L^{++}\,, \qquad\theta^{\pm}\,, \qquad\bar\theta^-\,, \qquad v^\pm\,, \qquad\bar v^\pm
$$
satisfying the quantum analogs of the first-class constraints
$$
d_\theta^+\,, \qquad\bar d_\theta^+\,, \qquad d_c^{++}\,, \qquad \bar d_c^{++}\,, \qquad d_c^{+-}\,, \qquad d_c^0\,,
$$
where the harmonics $v$ and $\bar v$ satisfy the kinematic conditions (5). The quantum analogs of the first-class
constraints are readily constructed by means of differentiation operators with allowance for their role as generators of
supertranslations and Lorentz transformations in the Borel subgroup. Thus, the quantum analogs of the first-class Fermi
constraints $d_\theta^+$ and $\bar d_\theta^+$ are realized as differentiations with respect to the corresponding isotropic coordinates:
$$
D^+\equiv \hat d^+| = \frac{\partial}{\partial\theta^-}\,, \qquad
\bar D^+\equiv \hat{\bar d}^+|\equiv   \frac{\partial}{\partial\bar\theta^-} + 2i\theta^- \frac{\partial}{\partial x_R^{--}} \,.
$$
A wave function $\Psi=\Psi_L$ that satisfies these constraints,
$$
\frac{\partial\Psi}{\partial\theta^-}=\frac{\partial\Psi}{\partial\bar\theta^-} =0\,,
$$
does not depend on $\theta^{-}$ and $\bar\theta^{-}$.
It is convenient to express the harmonic constraints in terms of the operators of the
corresponding infinitesimal transformations of scalar functions of the harmonic variables by setting
\begin{equation}
\begin{array}{l}
D^{++}\equiv i\hat d^{++}=v^+ {\displaystyle\frac{\partial}{\partial v^-}}\,,\qquad
\bar D^{++}\equiv \hat{\bar d}^{++}= \bar v^+ {\displaystyle\frac{\partial}{\partial \bar v^-}}\,, \\ [6pt]
D^{+-}\equiv i\hat d^{+-}={\displaystyle\sum_\pm}(\pm)
\left(v^\pm{\displaystyle\frac{\partial}{\partial v^\pm}}+\bar v^\pm{\displaystyle\frac{\partial}{\partial \bar v^\pm}}\right)\,,
 \\ [6pt]
D^{0}\equiv i\hat d^{0}={\displaystyle\sum_\pm}(\pm)
\left(v^\pm{\displaystyle\frac{\partial}{\partial v^\pm}}-\bar v^\pm{\displaystyle\frac{\partial}{\partial \bar v^\pm}}\right)\,.
\end{array}
\end{equation}
Then the wave function $\Psi=\Psi_L$, which depends on the isotropic coordinates $x_L^{++}$, $\theta^+$ and the harmonics $v^\pm$, $\bar v^\pm$, satisfies the
following equations (is subject to the quantum analogs of the corresponding first-class constraints):
\begin{equation}
D^{++}\Psi^{c,s}= \bar D^{++}\Psi^{c,s}=0\,,
\end{equation}
\begin{equation}
\left(D^{+-}+2x_L^{++}\frac{\partial}{\partial x_L^{++}} +\theta^{+}\frac{\partial}{\partial \theta^{+}} -2c\right)\Psi^{c,s}= 0\,,
\end{equation}
\begin{equation}
\left(D^{0} +\theta^{+}\frac{\partial}{\partial \theta^{+}} -2s\right)\Psi^{c,s}= 0\,.
\end{equation}
The complex ordering constants $c$ and $s$ -- the conformal and spin weights -- arise because of the ambiguity of the quantum
expressions associated with the constraints $d_c^{+-}$ and $d_c^{0}$ which contain products of noncommuting variables (coordinates and
their conjugate momenta). A wave function satisfying Eqs. (25) and (26) is equipped with the indices $c$ and $s$, which indicate
an arbitrary choice of the ordering constants. These must be determined by additional conditions of a physical kind. In
particular, from the requirement that the wave function be single valued it obviously follows that the ordering constant $2s$ is
an integer:
\begin{equation}
2s \in \mathbb{Z}\,.
\end{equation}

Note that the most general expression associated with the product $ab$ of non(anti)commuting fundamental quantities $a$ and $b$ has in the case of linear quantization the form
$$
ab\quad \leftrightarrow \quad \frac12\big\{\hat a ,\hat b \big] +k\big[\hat a ,\hat b \big\}\,,\qquad k\in \mathbb{C}\,.
$$
Here the pair of a curly and a square bracket means, depending on the order, the graded anticommutator, $\big\{ , \big]$, or the
commutator: $\big[ , \big\}$; $k$ is an ordering constant that is also associated with the choice of the fundamental classical quantities with
which the quantum operators are associated. To the values $k=0,1/2$ and $-1/2$ there correspond the Weyl, $ab$, and $ba$
orderings.

It is readily verified that Eqs. (24)--(26) are consistent for arbitrary values of the ordering constants $c$ and $s$.

Equations (24)~(26) can be considered in the central chiral basis (in the original variables
$x_L^{\mu}$, $\theta^{\alpha}$,  $v^\pm$, $\bar{v}^\pm$), where in the
differential operators of these equations there are no terms associated with the superspace coordinates, and the conformal and
spin weights keep their previous values:
\begin{equation}
D^{++}\Psi^{c,s}= \bar D^{++}\Psi^{c,s}=0\,,
\end{equation}
\begin{equation}
\left(D^{+-} -2c\right)\Psi^{c,s}= 0\,,
\end{equation}
\begin{equation}
\left(D^{0}  -2s\right)\Psi^{c,s}= 0\,.
\end{equation}
Here it is necessary to make the superfield $\Psi^{c,s}$ satisfy the harmonicity conditions that ensure its dependence on the coordinates
of the chiral superspace $$
x_L^{\mu}=x^{\mu}+i\theta\sigma\bar\theta
$$
and $\theta^\alpha$, namely, through $x_L^{++}$ and $\theta^+$:
\begin{equation}
u^{++} \partial_L\,\Psi^{c,s}=u^{\pm\mp} \partial_L\,\Psi^{c,s}=0\,,
\end{equation}
\begin{equation}
 v^+D_L\,\Psi^{c,s}={\bar D}_L\,\Psi^{c,s}=0\,,
\end{equation}
where
$$
D_L=\partial/\partial\theta-2i(\hat\partial_L\bar\theta)\,,\qquad
\bar D_L=\partial/\partial\bar\theta
$$
are covariant spinor derivatives in the chiral basis. The first of the conditions (31) follows from the remaining conditions and
Eq. (28).

Because of the completeness of the harmonic basis in the spinor space, the conditions (31) can also be expressed in the
equivalent more compact form
\begin{equation}
v^+ \hat\partial_L\,\Psi^{c,s}= \hat\partial_L\bar v^+\,\Psi^{c,s}=0\,.
\end{equation}
These conditions commute with Eq. (28), whereas the conditions (31) do not.

Of course, the sets of differential operators in (28)--(32) and (28)--(30), (32)--(33) are closed with respect to the operation
of commutation.

Note that Eqs. (28)--(30) reduce the number of essential harmonic variables to one complex parameter, which can be taken
to be the complex coordinate of the point of the celestial sphere to which the ray associated with the propagation of the massless
superparticle points. Thus, the compact harmonic space is the complex projective line $\mathbf{C}\mathbf{P}^1$.

Indeed, in accordance with (29)--(30) the harmonic superfield $\Psi^{c,s}(x_L^{++},\theta^+,v^\pm)$ is an eigenvector of the generators of
transformations in the parabolic subgroup (stationary group of the complex projective line) -- the nilpotent and boost [or
$SO{\uparrow}(1,1)$] transformations and rotations [or $SO(2)$] with eigenvalues 0, $2c$, $2s$, respectively. Therefore, the harmonic
superfield is invariant with respect to gauge transformations in the nilpotent subgroup:
\begin{equation}
\Psi^{c,s}(v^+,v^-+\xi v^+)=\Psi^{c,s}(v^+,v^-)\,,\qquad \forall \xi \in \mathbb{C}\,,
\end{equation}
and is homogeneous, with homogeneity degrees $2c$ and $2s$, with respect to the diagonal $SO{\uparrow}(1,1)$ and $SO(2)$ transformations,
which act multiplicatively on the harmonic indices:
\begin{equation}
\Psi^{c,s}(e^{\pm a}v^\pm)=e^{2ca}\Psi^{c,s}(v^\pm)\,,\qquad a \in \mathbb{R}\,,
\end{equation}
\begin{equation}
\Psi^{c,s}(e^{\pm i\phi}v^\pm)=e^{2is\phi}\Psi^{c,s}(v^\pm)\,,\qquad \phi \in \mathbb{R}\,.
\end{equation}
Because of the unimodularity of the $2{\times}2$ matrices with columns $v^+$ and $v^-$, i.e., because $v\in SL(2,\mathbb{C})$, or, equivalently, because of the kinematic constraints (5), we can regard as independent only one complex coordinate of the spinor $v^-$, the one
corresponding to the component collinear with $v^+$. However, this last can be arbitrarily changed by a gauge transformation
(34) in the nilpotent subgroup. This makes it possible to regard $\Psi^{c,s}$ as a function of just two complex variables -- the
coordinates of the spinor v +, which do not vanish simultaneously, i.e.,
$$
v^+ \,\in \,\,\stackrel {o} {\mathbb{{C}}^2}\equiv \mathbb{C}^2\setminus\{0\}\,,
$$
this being a direct consequence of the kinematic condition $v^-v^+=1$. We give the explicit form of the gauge transformation that
eliminates $v^-$:
$$
\xi=(v^-\sigma_0\bar v^+)/(v^+\sigma_0\bar v^+)
$$
gives
$$
v^-=-(\sigma_0\bar v^+)/(v^+\sigma_0\bar v^+)\,.
$$
To obtain the general solution of the unimodularity condition, it is necessary to add a term proportional to $v^+$ on the right-hand
side of the last equation.

Equations (35) and (36) now mean that the wave function as a function of $v^+$ is homogeneous with index $\chi=(n_1,n_2)$,
where $n_{1}=c+ s+1$, $n_{2}=c- s+1$:
$$
\Psi^{c,s}(av^\pm)=a^{n_1-1}\bar a^{n_2-1}\Psi^{c,s}(v^\pm)
$$
for all nonvanishing complex $a$. One sometimes speaks of a bidegree $(\nu_1,\nu_2)$, where $\nu_{1,2}=n_{1,2}-1$. When considering the
corresponding representations of the Lorentz group, it is convenient to use in place of $\chi$ the pair $[l_0,l_1]$, where $l_0=s$, $l_1=c+1$.

As is well known [19], homogeneous infinitely differentiable functions of two complex variables correspond bijectively
to infinitely differentiable (together with the inversion
$$
\tilde\psi(z)=z^{n_1-1}\bar z^{n_2-1}\psi(1/z)
$$
of the same bidegree) functions
$$
\psi(z)=\Psi^{c,s}(z,1)\,,
$$
where $z=v^+_1/v^+_2$, of one complex variable. Constructing for $v^+=(z,1)$ a
complex isotropic tetrad and using the standard connection $z=e^{i\phi}\mathrm{ctg}\, \theta/2$
of the angle coordinates $\theta, \phi$ on the sphere with
coordinate $z$ on the complex projective line (realized as the equatorial plane augmented by the point at infinity), it is easy to
show that the functions $\psi(z)$ and their inversions $\tilde\psi(z)$ correspond to functions on the sphere under a cartographic mapping that
is regular at its south and north poles, respectively.

In the locally convex vector topological space of homogeneous infinitely differentiable functions of homogeneity degree
$\chi$ on the complex affine plane $\stackrel{o}{\mathbb{C}^2}$
there is realized an infinite-dimensional representation of the proper Lorentz group $SL(2,\mathbb{C})$,
which in general is irreducible. Only at integer points, i.e., when $n_1$ and $n_2$  are nonvanishing integers of the same sign or,
equivalently, $c$ and $s$ are both integers or half-integers satisfying the condition $|c+1|>|s|$, does there exist in $D_{\chi}$ an invariant
subspace and the representation is not completely reducible. At positive integer points ($n_{1,2}>0$  and we write $\chi>0$), the
invariant subspace $E_{\chi}$ is a finite-dimensional space of homogeneous polynomials of dimension $n_1 n_2$, in which there is realized
a nonunitary (for $n_1+n_2>2$) irreducible finite-dimensional representation; the representation in the factor space $D_{\chi}/E_{\chi}$ is
equivalent to an infinite-dimensional irreducible representation in the space
$$
D_{(-n_1,n_2)}\simeq D_{(n_1,-n_2)}\,.
$$
At the negative integer points, the invariant subspace is an infinite-dimensional space $F_{\chi}$ of functions with zero moments: In
the realization of the space $D_{\chi}$ on the sphere this means that
$$
\int z^{j_1}\bar z^{j_2}dz d\bar z =0 \quad\mbox{for}\quad 0\leq j_{1,2}\leq |n_{1,2}|\,.
$$
The representation in the space $F_{\chi}$ is equivalent to an irreducible infinite-dimensional representation in the space
$$
D_{(-n_1,n_2)}\simeq D_{(n_1,-n_2)}\,,
$$
the factor space $D_{\chi}/F_{\chi}$ is finite dimensional, and the representation in it is equivalent to a representation in the space $E_{-\chi}$ of homogeneous polynomials.

Nontrivial unitary representations (necessarily infinite dimensional) are obtained when either $c$ is purely imaginary
(principal series) or when $s=0$ and the real $c$ satisfies the condition $0<|c+1|<1$ (complementary series). Incidentally,
nonunitarity of a representation of the Lorentz group acting in the space of indices of a (super)field does not prevent us from
associating with this field (super)particles described by unitary representations of the (super) Poincar\'{e} group.

Thus, as a result of the first operator quantization of the harmonic superparticle we always obtain an infinite-component
superfield -- a homogeneous (of index $\chi$) infinitely differentiable function
$$
\Psi^{c,s}(x_L^{++},\theta^+,v^+)
$$
of the spinor $v^+$ (which ranges
over the complex affine plane $\stackrel{o}{\mathbb{C}^2}$) that must satisfy the harmonicity conditions.
It is only at integer points that one can separate
a finite-component superfield, on the index of which there is realized a nonunitary (for $n_1+n_2\neq 0$) finite-dimensional
representation. When $n_1=n_2=1$, a field is separated with a trivial, and therefore unitary representation on the index. Such
a field does not depend explicitly on $v^+$, this being admissible by virtue of the compactness of the space of harmonics.

We recall that the obtained superfield satisfies the condition of masslessness (21), whose solution, which does not depend
on $x_R^{--}$, satisfies the constraints (28)--(30) and also the chirality and harmonicity conditions (32)--(33).

\section{Analysis of the spectrum}

The harmonic superfield
$$
\Psi^{c,s}(x,\theta,v^+)\equiv \Psi_\chi(x,\theta,v^+)\,,
$$
obtained as a result of the first quantization of the harmonic superparticle, transforms in accordance with an infinite-dimensional
representation of the quantum-mechanical Lorentz group in the index space parametrized by the spinor $v^+$. The spectrum of
states described by it (irreducible unitary representations of the Poincar6 supergroup) is determined by the values of the Casimir
operators. Of course, by construction these operators are massless, and we shall show below, having calculated the
Pauli--Luba\'{n}ski vector, that the corresponding particles necessarily have finite spin (helicity $-s$).

It is clear that because of the harmonic gauge properties (34)--(36) of the harmonic superfield the prescription of canonical
quantization cannot be directly applied. It would be natural to establish a linear connection between the equivalence classes of
the harmonic gauge fields and the ordinary, nongauge fields that admit standard canonical quantization. It turns out that such
a correspondence is bijective and is uniquely determined if it is required to be translationally invariant. To construct, as usual,
this correspondence by the method of Cartan forms, we find a quasiinvariant measure $\omega\wedge\bar\omega$ in the complex affine plane of the
spinor variable $v^+$ (containing the space of harmonics $\mathbf{C}\mathbf{P}^1$). Here
$$
\omega=\frac14\,dv^+\wedge dv^+\,.
$$
It is convenient to represent an integral around a piecewise smooth contour surrounding the origin in the complex plane (i.e.,
over the harmonic space) by means of an arbitrary function
$$
H(v^+) \in D(\stackrel{o}{\mathbb{C}^2})
$$
such that
\begin{equation}
\int_{\stackrel{o}{\mathbf{\mathbb{C}}^2}}\,H(av^+)\frac{|dad\bar a|}{|a|^2}=1 \,,\qquad
\forall\, v^+\in\stackrel{o}{\mathbb{C}^2}\,,\qquad
|dad\bar a|=da\wedge d\bar a/2i\,,
\end{equation}
in the form of an integral over the complete plane $\stackrel{o}{\mathbb{C}^2}$ with respect to the measure
$$
[\omega\wedge\bar\omega](v^+)\equiv\omega\wedge\bar\omega\cdot H(v^+)\,.
$$
Then the residue of the homogeneous function of bidegree $(-2,-2)$ does not depend on the choice of $H$ [37].

It is obvious that the superfield determined by the expression
\begin{equation}
  \Phi_{-\chi}^{(+)}(\zeta,x,\theta)=\int[\omega\wedge\bar\omega](v^+)
  K_{-\chi}(\zeta v^+)\Psi_{\chi}^{(+)}(x,\theta,v^+)  \,,
\end{equation}
is gauge covariant and homogeneous of index $-\chi$ with respect to the spinor $\zeta$ and does not depend on the choice of the
representative of the harmonic gauge class (on the gauge fixing for the nilpotent subgroup). Here the kernel of the
transformation
\begin{equation}
  K_\chi(z)=|z|^{2c}\exp(2is\arg z)/ \Gamma(c+|s|+1)
\end{equation}
is a generalized function of one complex variable $z$ that is homogeneous with index $\chi$ and analytic with respect to $c$ for fixed
(half-)integral $s$. It is known [19] that at noninteger points $\chi$ an operator that commutes with the representation of the group
$SL(2,\mathbb{C})$ is uniquely determined and vice versa: The result of successive application of two transformations with kernels
$K_{-\chi}$ and $K_\chi$ is a multiple of the identity transformation with coefficient
$$
\pi^2(-1)^{2s}/\Gamma(c+|s|+1)\Gamma(-c+|s|)\,.
$$

At integer points, the kernel of the integral transformation (38)--(39) (a homogeneous generalized function) is not uniquely
defined; this difficulty can be eliminated by choosing a definite analytic continuation with respect to the homogeneity index $\chi$.
Thus, several mappings are obtained, among which the most important maps the space $D_\chi$ to an invariant subspace of the space
$D_{-\chi}$, becoming degenerate on an invariant subspace of the space $D_\chi$. The mapping from the finite-dimensional invariant
subspace $E_\chi$ need not be considered, since its image will also be finite dimensional [19], and the state spaces of real particles
are infinite dimensional. If we are interested in only massless fields that satisfy the microcausality condition, it is also
unnecessary to consider the remaining mappings, since their image will not have integer positive homogeneity index [19].

In any of the considered cases, the kernel is a generalized function of the contraction $\zeta v^+$. Therefore, the conditions of
harmonicity (32)--(33) of the superfield $\Psi_\chi$ are translated into equations of motion [see (51) below] for the superfield $\Phi_{-\chi}$, the
operators of which are obtained from the operators in the harmonicity conditions (32)--(33) by the simple substitution
$$
v^+\quad\rightarrow\quad \partial/\partial\zeta\,.
$$

The harmonic superfield
$$
\Psi^{c,s}(x_L^{++},\theta^+,v^+,\bar v^+)\,,
$$
obtained as a result of the first quantization of the harmonic superparticle, describes a massless [see (20)--(21)] superparticle
of finite superspin (superhelicity). This last result is a consequence of the twistor representation
\begin{equation}
p_{\alpha\dot\alpha}=p_L^{--}v_{\alpha}^{+}\bar v_{\dot\alpha}^{+}
\end{equation}
for the 4-momentum of the superparticle (see, for example, [17]). In the general case, a massless particle can have infinite spin
(see, for example, [38]).

We show that the harmonic superparticle has finite superspin by explicit calculation of the Casimir operators of the
Poincar6 supergroup realized on harmonic superfields. We represent the generators of the Lorentz group in the form
$$
M_{\alpha\dot\alpha\beta\dot\beta} =2i\left(\epsilon_{\dot\alpha\dot\beta} M_{\alpha\beta} +
\epsilon_{\alpha\beta} \bar M_{\dot\alpha\dot\beta} \right)\,,
$$
where the symmetric spinors $M_{\alpha\beta}$ and $\bar M_{\dot\alpha\dot\beta}$ are the (anti)self-dual parts of the generators,
\begin{equation}
M_{\alpha\beta}=v^+_{(\alpha}\frac{\partial}{\partial v^{+\beta)}} \,,\qquad
\bar M_{\dot\alpha\dot\beta}= \bar v^+_{(\dot\alpha}\frac{\partial}{\partial \bar v^{+\dot\beta)}}
\end{equation}
In accordance with (40)
\begin{equation}
P_{\alpha\dot\alpha} = 2i v^+_{\alpha}\bar v^+_{\dot\alpha}\frac{\partial}{\partial x^{++}_L} \,,
\end{equation}
and from (17)--(18) we obtain
\begin{equation}
Q_{\alpha}=v^+_{\alpha}\frac{\partial}{\partial \theta^{+}} \,,\qquad
\bar Q_{\dot\alpha} = 4i \bar v^+_{\dot\alpha}\theta^{+}\frac{\partial}{\partial x^{++}_L}\,.
\end{equation}

The masslessness ($p^2=0$) for the representation (42) is obvious ($(v^{+})^2=(\bar v^{+})^2=0$).

The Pauli--Luba\'{n}ski pseudovector
$$
\omega_{\alpha\dot\alpha} =P_{\alpha}{}^{\dot\beta} \bar M_{\dot\beta\dot\alpha} -
P_{\dot\alpha}{}^{\beta} M_{\beta\alpha}
$$
has the supersymmetric generalization
$$
W_{\alpha\dot\alpha}=\omega_{\alpha\dot\alpha} +\frac14\,\left[ Q_{\alpha},\bar Q_{\dot\alpha}\right]\,,
$$
which possesses simple commutation properties with the supercharges:
$$
\left[ W_{\alpha\dot\alpha},Q_{\beta}\right]= \frac12\,P_{\alpha\dot\alpha}Q_{\beta}\,.
$$
In the realization (41)--(43), the vector $W_{\alpha\dot\alpha}$ is proportional to the 4-momentum:
$$
W_{\alpha\dot\alpha}=-\frac12\left( D^0+ \left[ \theta^+,\frac{\partial}{\partial\theta^+}\right]\right)P_{\alpha\dot\alpha}\,.
$$
Therefore, the harmonic superfield
$$\Psi^{c,s}(x_L^{++},\theta^+,v^+,\bar v^+)$$
describes a superparticle of finite superspin. The superhelicity operator
$$
\Lambda=-\frac12\, D_c^0=-\frac12\left( D^0+ \theta^+\frac{\partial}{\partial\theta^+}\right)\,,
$$
which commutes with all generators of the Poincar\'{e} supergroup, is found, as usual, from the equation
$$
W_{\alpha\dot\alpha} -AP_{\alpha\dot\alpha}=\Lambda P_{\alpha\dot\alpha}\,,
$$
where
$$
A= \frac12\,\theta^+\frac{\partial}{\partial\theta^+}
$$
is the generator of phase transformations of the Grassmann coordinates.

Thus, the harmonic superfield with spin weight $s$  describes massless superparticles of superhelicity
$$
\Lambda=-s\,.
$$
The components of the superfield
$$
\Psi^{c,s}(x_L^{++},\theta^+,v^+,\bar v^+)=
\psi_0^{c,s}(x_L^{++},v^+,\bar v^+)+
\theta^+\psi_1^{c-\frac12,s-\frac12}(x_L^{++},v^+,\bar v^+)
$$
-- the harmonic fields $\psi_0$ and $\psi_1$ -- describe massless particles
of superhelicities $\lambda_0=-s$ and $\lambda_1=-s+\frac12$, respectively.

The conformal weight $c$ of the harmonic superfield $\Psi^{c,s}$ determines the particular method of field description of the
superparticle of superhelicity $-s$ and is not reflected in the spectrum of states of the free particle described by the superfield.

In the twistor approach, positivity of the energy is automatically ensured. In a study using harmonics of an ordinary (not
super) particle, the sign of the energy is equal to the sign of the isotropic component $p^{--}$ of the momentum and must be
postulated. In the case of the superparticle, positivity of the energy is ensured by supersymmetry.

\section{Massless nongauge fields}

By definition, a Lorentz-covariant (super)field transforms in accordance with the equation
\begin{equation}
  U(A)\Phi_i(x,\theta)U(A)^{-1}=(T(A^{-1})\Phi)_i(\Lambda(A)x,\theta A^{-1}),
\end{equation}
where $U(A)$ is the unitary operator of a representation in the Hilbert state space of the element $A$ of the quantum-mechanical
Lorentz group $SL(2,\mathbb{C})$, $T(A)$ is some representation of the group $SL(2,\mathbb{C})$ that acts only on the index
$i$,  $x^\prime=\Lambda(A)x$, $\hat x^\prime=A\hat x A^+$.
One and the same (super)particle described by an irreducible representation of the Poincar6 (super)group can be described by
different types of fields characterized by different representations $T(A)$.
To have the possibility of describing fields on whose
index an arbitrary (not necessarily finite-dimensional) irreducible representation of the Lorentz group is realized, one introduces
an ``index spinor'' $\zeta\neq 0$, which ranges over the complex affine plane $\stackrel {o} {{\mathbb{C}}^2}$ [38].
Then the finite-component fields with irreducible
representation $(j,k)$ on the index stand in a one-to-one relation with the homogeneous polynomials $\Phi(\zeta;x,\theta)$ of homogeneity
degree $(2j+1,2k+1)$ in $\zeta$ obtained by contracting the spinor indices of the superfield $\Phi_i(x,\theta)$, which together replace the
selective index $i$, with the index spinor $\zeta$ in the case of an undotted index and with its conjugate $\bar\zeta$ in the case of a dotted index.
In the language of the index spinor, the transformation law (44) of the superfield takes the form
$$
  U(A)\Phi(\zeta;x,\theta)U(A)^{-1}=
  \Phi(\zeta A^{-1};\Lambda(A)x,\theta A^{-1}).
$$
All operator-irreducible representations of the group $SL(2,\mathbb{C})$ can be realized on suitable (convex topological vector) spaces of
homogeneous functions of a pair of complex variables $\zeta=(\zeta^1,\zeta^2)\in\stackrel {o} {{\mathbb{C}}^2}$ [19],
and therefore the superfield $\Phi_i(x,\theta)$ with arbitrary
irreducible representation of the Lorentz group on the index $i$ can be regarded as a superfield $\Phi(\zeta;x,\theta)$ that is a homogeneous
function of appropriate homogeneity degree of the index spinor $\zeta$; as usual, homogeneity of bidegree $\chi=(n_1,n_2)$ means that
$$
\Phi(a\zeta)=a^{n_1-1}\bar a^{n_2-1}\Phi(\zeta)\,, \qquad \forall a\in
\stackrel {o} {{\mathbf{C}}^2}\,,
$$
where the numbers $n_1$ and $n_2$ are complex and their difference is an integer. Then the infinitesimal operators of the index
representation are realized as
$$
M_{\mu\nu}=i\left(\zeta\sigma_{\mu\nu}\frac{\partial}{\partial\zeta}+\bar\zeta\tilde\sigma_{\mu\nu}\frac{\partial}{\partial\bar\zeta} \right)\,,
$$
where $\sigma_{\mu\nu}=\sigma_{[\mu}\tilde\sigma_{\nu]}/2$, $\tilde\sigma_{\mu\nu}=\tilde\sigma_{[\mu}\sigma_{\nu]}/2$.

If we not distinguish equivalent representations, then without loss of generality we can restrict consideration to just the
noninteger points, when the representations $\chi$ and $-\chi$ are equivalent, and positive integer points and choose functions in the
invariant space of homogeneous polynomials.

We restrict consideration to ordinary free fields that satisfy the massless Klein--Gordon equation, stipulating details
associated with the supersymmetry as necessary. The second quantization of such fields is done in the framework of the usual
formalism. The annihilation field for a particle with helicity $\lambda$ is
$$
  \varphi_\chi^{(+)} (\zeta,x)=(2\pi)^{-\frac 32}
  \int\,d^4 p\delta_+(p^2)e^{ipx}u_\chi(\zeta,p;\lambda)a(\vec p,\lambda) ,
$$
where $\delta_+(p^2)\equiv\theta(p^0)\delta(p^2)$, $u_\chi$ is the Wigner wave function in the momentum representation,
$a(\vec p,\lambda)$ is the annihilation
operator, and $\chi$ is the bidegree that determines the irreducible representation with respect to which the field index transforms.
In the creation field $\varphi_\chi^{(-)}(\zeta,x)$, which transforms in exactly the same way under Poincar\'{e} transformations, the product
$$
e^{-ipx}v_\chi(\zeta,p;\lambda^\prime)b^+(\vec p,\lambda^\prime)
$$
with the operator $b^+(\vec p,\lambda^\prime)$ of creation of an antiparticle
of helicity $\lambda^\prime$ and wave function $v_\chi(\zeta,p;\lambda^\prime)$ is integrated with respect
to the same measure. The creation and annihilation operators of particles and antiparticles of opposite helicities transform with
respect to one-dimensional contragredient representations of the Lorentz group as follows:
$$
  U(A)a(\vec p,\lambda)U(A)^{-1}=
  e^{-i\lambda\Theta(R)}a(\vec p^\prime,\lambda)\,,
$$
where $R$ is a transformation of the little group $E(2)$ of the standard momentum $\stackrel {o}{p}=(k,0,0,k)$,$k>0$, corresponding to the
transformation $A$ of $SL(2,\mathbb{C})$ for fixed Wigner operator $B_p$:
$$
R=B_{p^\prime}^{-1}AB_p\,.
$$
The rotation angle $\Theta(R)$ in the Euclidean plane is determine by the equation
$$
exp(i\Theta(R)/2)\hat\sigma=
\hat\sigma R\hat\sigma\,,
$$
where $\hat\sigma=(1+\sigma_3)/2$.
We determine the Wigner operator in the form of the product of a rotation $R(\vec p)$ that carries the unit vector
of the third axis into the momentum unit vector $\vec p/|\vec p|$ and a purely Lorentz transformation along the third axis
$H(|\vec p|)=exp(\frac 12\ln\frac{|\vec p|}k\sigma_3)$:
$$
B_p=R(\vec p)H(|\vec p|)\,.
$$

A wave function of arbitrary momentum is obtained from the wave function of the standard momentum by a transformation
of only the index spinor:
$$
  u_\chi(\zeta,p;\lambda)=u_\chi(\zeta B_p,\stackrel {o} {p};\lambda)\,.
$$
From this we can readily obtain the transformation law of the wave function of the standard momentum:
\begin{equation}
  (T(R)u_\chi)(\zeta,\stackrel {o} {p};\lambda)  =
  e^{i\lambda\Theta(R)}u_\chi(\zeta,\stackrel {o} {p};\lambda).
\end{equation}

The little group is generated by the components of the Pauli--Luba\'{n}ski vector
$$
w^\lambda=\frac 12\epsilon^{\lambda\mu\nu\rho}M_{\mu\nu}P_\rho=
k(J_3,L_1,L_2,J_3)
$$
in the system of the standard momentum by the generators of the algebra of the group of motions $E(2)$ of the two-dimensional
Euclidean plane. Here
$$
J_j=\frac 12\epsilon_{jkl}M_{kl}
$$
are the generators of three-dimensional rotations,
$$
L_{1,2}=J_{1,2}\mp K_{2,1}\,,
$$
where $K_j=M_{0j}$ are
boosts. In the neighborhood of the identity of the little group, we have, if $\theta,y_1,y_2$ are its parameters,
$\Theta(R)(\theta,y_1,y_2)\simeq\theta$.
Therefore, for the infinitesimal transformations in (45) we have [18]
\begin{equation}
(A_3+B_3)u_\chi(\zeta,\stackrel {o} {p};\lambda)=
\lambda u_\chi(\zeta,\stackrel {o} {p};\lambda)\,,
\end{equation}
\begin{equation}
(A_1+iA_2)L_1u_\chi(\zeta,\stackrel {o} {p};\lambda)=0\,,
\end{equation}
\begin{equation}
(B_1-iB_2)L_1u_\chi(\zeta,\stackrel {o} {p};\lambda)=0\,.
\end{equation}
Here the components of the 3-vectors $A_i$ and $B_i$ form the Lie algebras of the groups $SU(2)$:
$$
[A_i,A_j]=i\varepsilon_{ijk}A_k\,,\quad [A_i,B_j]=0\,,\quad \mathbf{J}=\mathbf{A}+\mathbf{B}\,,
\quad \mathbf{K}=-i(\mathbf{A}-\mathbf{B})\,;
$$
$$
\mathbf{A}^2=\frac{\nu_1}{2}\left(\frac{\nu_1}{2}+1\right)\,,\quad \mathbf{B}^2=\frac{\nu_2}{2}\left(\frac{\nu_2}{2}+1\right)\,,\quad
A_j=\frac 12\zeta\sigma_j\frac\partial{\partial\zeta}\,,\quad B_j=-\frac 12\bar\zeta\sigma_j^T\frac\partial{\partial\bar\zeta}\,.
$$
The freedom in the choice of the standard-momentum system makes it possible to use when necessary nilpotent transformations
and assume without loss of generality that $\zeta^1\neq 0$. Then it follows from (47)--(48) that the wave function does not depend on
$\zeta^2$. Therefore, $A_3$ and $B_3$ when they act on the wave function are proportional to the operators that determine the homogeneity
index:
$$
A_3u_\chi(\zeta,\stackrel {o} {p};\lambda)=
\frac{\nu_1}{2}\,u_\chi(\zeta,\stackrel {o} {p};\lambda)\,,
$$
$$
B_3 u_\chi(\zeta,\stackrel {o} {p};\lambda)=
-\frac{\nu_2}{2}\,u_\chi(\zeta,\stackrel {o} {p};\lambda)\,,
$$
Therefore, the helicity is $\lambda=(\nu_1 -\nu_2)/2=s$.

Thus, the spin weight $s$ of a nongauge field describing massless particles of finite spin is necessarily equal to the helicity
$\lambda$ of the particles:
$$
s=\lambda\,.
$$
For finite-component fields with $c\geq|s|$, this result was obtained by Weinberg [18].

Our arguments also show that the wave function in the standard-momentum system is a homogeneous function of index
$\chi=(n_1,n_2)$ of the first component $\zeta^1$ of the index spinor $\zeta$. Accordingly, the wave function of an arbitrary momentum has the
form
$$
  u_\chi(\zeta,p;\lambda)=
  {[\zeta B_p]}^{c+s} {\overline{[\zeta B_p]}}
  ^{c-s} / \Gamma(c+|s|+1) ,
$$
where by definition $[\zeta B_p]\equiv(\zeta B_p)^1$ The introduction of the gamma function makes it possible to assume that the right-hand side
is an analytic function of conformal weight $c$ for fixed spin $s$. Since we consider free fields, the possible constant factor is
omitted.

Thus, there are four conditions that determine the dependence of the wave function on the index spinor $\zeta$: Two must ensure
that it depends only on the first component in the standard-momentum system, and two must determine the homogeneity degree
(one the spin weight $s$ or helicity $\lambda=s$, the other the conformal weight, i.e., the choice of the method of field description).
When ordinary fields with indices are considered, the spin and conformal weights are fixed by specifying a set of spinor and/or
4-vector field indices, and of the two remaining conditions one may not be needed if the field has indices of only one type, for
example, only undotted.

In manifestly Lorentz-covariant form for a wave function of definite homogeneity degree, the conditions that determine
the dependence on the index spinor (the equation of motion) have the form
\begin{equation}
  \frac\partial{\partial\zeta}\hat p \, u_\chi(\zeta,p;\lambda)=0\,,\qquad
  \hat p\frac\partial{\partial\bar\zeta} \, u_\chi(\zeta,p;\lambda)=0.
\end{equation}
From this we readily obtain eigenvalue equations for the operators of the helicity and boosts along the momentum:
$$
\left(\mathbf{p}\mathbf{J}-\lambda|\mathbf{p}| \right) u_\chi(\zeta,p;\lambda)=0\,,\qquad
\left(\mathbf{p}\mathbf{K}-ic|\mathbf{p}| \right) u_\chi(\zeta,p;\lambda)=0.
$$
For a finite-component field with (un)dotted indices, the first of these, which is equivalent to (49), was considered together with
the Klein--Gordon equation in [18] as the field equation of motion.

Since the antiparticle wave function is
$$
  v_\chi(\zeta,p;\lambda^\prime)=
  \overline{u_{\bar\chi}(\zeta,p;-\lambda)} \,,
$$
where $\bar\chi\equiv(\bar c,-s)$, its helicity
$$
\lambda^\prime=-s
$$
is opposite to that of the particle's helicity.

The complete quantum field $\varphi_\chi(\zeta,x)$ can be expressed as a combination of positive- and negative-frequency parts with,
in general, arbitrary complex weights $\xi_\chi$ and $\tilde\xi_\chi$:
\begin{equation}
  \varphi_\chi(\zeta,x)=\xi_\chi\varphi_\chi^{(+)}(\zeta,x)+
  \tilde\xi_\chi\varphi_\chi^{(-)}(\zeta,x)\,.
\end{equation}
Of course, the phases of the weight factors can be eliminated by a redefinition of the particle and antiparticle creation operators.
By virtue of (49), the field (50) satisfies in addition to the massless Klein--Gordon equation the equations of motion
\begin{equation}
  \frac\partial{\partial\zeta} \hat\partial \, \varphi_\chi(\zeta,x)=0\,,\qquad
  \hat\partial\frac\partial{\partial\bar\zeta} \, \varphi_\chi(\zeta,x)=0
\,.
\end{equation}

To relate the spinor basis in the standard-momentum system to the spinor basis in the light-cone system adapted to the 4-
momentum of the particle, it is necessary to make a boost transformation in the direction of the standard momentum, carrying
its standard isotropic coordinate into a given one, a rotation, which matches the phase of the first component of the first of the
basis spinors with the phase of the first component of the harmonic spinor $v^+$, and a nilpotent transformation in the stability
group of the standard momentum that identifies the second basis spinor with the harmonic spinor $v^-(v^+)$. The composition
of these transformations gives a representation of the Wigner operator in the isotropic basis. To obtain a representation in the
ordinary basis, one can use the harmonic matrix $v$ as a bridge between the ordinary spinor basis and the isotropic basis. Then
for the Wigner operator we obtain the expression
$$
  B_p=\left(v_\alpha^+{\left(\frac{p^{--}}{2k}\right)}^{\frac 12}
  {\left(\frac{v_1^+}{\bar v_1^+}\right)}^{-\frac 12} \, ,  \,
  v_\alpha^-(v^+){\left(\frac{p^{--}}{2k}\right)}^{-\frac 12}
  {\left(\frac{v_1^+}{\bar v_1^+}\right)}^{\frac 12}\right),
$$
where we have omitted the in general arbitrary phase of the component of the first basis spinor of the standard-momentum
system. This makes it possible to write the Wigner wave function of the particle in the form
$$
  u_\chi(\zeta,p;\lambda)=K_\chi(\zeta v^+)
  \left(\frac{p^{--}}{2k}\right)^c\left(\frac{v_1^+}{\bar v_1^+}\right)^{-s}\,.
$$
In the isotropic basis adapted to the 4-momentum, the 4-momentum can be expressed in terms of the harmonic spinor by Eq.
(40). Then the Lorentz-invariant measure on the light cone is
$$
d^3\mathbf{p}/|2\mathbf{p}|=\frac{1}{2i} v^+dv^+\bar v^+d\bar v^+p^{--}dp^{--}\,,
$$
where the harmonic spinor describes a closed piecewise smooth surface in the complex affine plane that surrounds the origin
and intersects each straight line that passes through the origin at one point. The annihilation field of a particle of helicity $\lambda=s$
can be represented as the surface integral
$$
  \varphi_\chi^{(+)}(\zeta,x)=\frac 1{2i}
  \int v^+dv^+\bar v^+d\bar v^+ K_\chi(\zeta v^+)
  \psi_{-\chi}^{(+)}(x^{++},v^+,\bar v^+)
$$
over the described surface of the operator function
\begin{equation}
  \psi_{-\chi}^{(+)}(x^{++},v^+)=\frac 1{(2k)^c(2\pi)^{\frac 32}}
  \int_0^\infty\,dp^{--}e^{-\frac i2 p^{--}x^{++}}(p^{--})^{c+1}
  \left(\frac{v_1^+}{\bar v_1^+}\right)^{-s}a(\mathbf{p},s)\,,
\end{equation}
which is homogeneous of degree $-\chi$:
$$
  \psi_{-\chi}^{(+)}(|a|^2 x^{++},av^+)=|a|^{-2(c+2)}
  \exp(-2is\arg a)\psi_{-\chi}^{(+)}(x^{++},v^+,\bar v^+).
$$
Because of this, it is possible to rewrite the surface integral as an integral over the complete affine plane
$\stackrel{o}{\mathbb{C}^2}$, using the measure
$[\omega\wedge\bar\omega](v^+)$ (37)--(38):
\begin{equation}
  \varphi_{\chi}^{(+)}(\zeta,x)=\int[\omega\wedge\bar\omega](v^+)
  K_{\chi}(\zeta v^+)\psi_{-\chi}^{(+)}(x^{++},v^+,\bar v^+)  \,.
\end{equation}
A representation of the form (52) also holds for the creation field
$\varphi_{\chi}^{(-)}(\zeta,x)$ with corresponding substitution of the antiparticle
creation operator $b^+(\mathbf{p},-\lambda)$ in place of $a(\mathbf{p},\lambda)$
and replacement of the positive-frequency exponential by the negative-frequency
one, and there is also such a representation for the complete massless field $\varphi_{\chi}(\zeta,x)$. For integer values of the homogeneity
index $\chi$, the integral transformation in (53) has a kernel -- an invariant subspace in the space of homogeneous functions of the
harmonic spinor $v^+$ of index $-\chi$. Therefore, at integer points the massless field $\varphi_{\chi}(\zeta,x)$
is associated with an equivalence class
of harmonic fields with respect to the addition of an arbitrary homogeneous function belonging to the invariant subspace of
homogeneous functions of appropriate homogeneity index.

The expression (52) makes it possible to formulate a simple ansatz for the second quantization of the harmonic fields
$$
\psi_{-\chi}^{(+)}(x^{++},v^+,\bar v^+)\,.
$$
It is necessary to perform a Fourier transformation of the harmonic field with respect to the isotropic
coordinate $x^{++}$ and to separate in the integrand the $c+1$ and $-2s$ powers of the momentum $p^{--}$ and the argument of the first
component of the harmonic spinor $v^+$, which ensure the correct conformal and spin weights. The remaining part of the
integrand can be identified, up to the normalization factor $(2k)^{-c}(2\pi)^{-3/2}$, with the operators of annihilation
$a(\mathbf{p},\lambda)$ and creation
$b^+(\mathbf{p},\tilde\lambda)$ of physical particles and antiparticles of helicities $\lambda=s$ and $\tilde\lambda=-s$ in the positive- and negative-frequency parts, respectively.

\section{Commutation function and microcausality}

It is well known that in the case of infinite-component fields describing particles of infinite spin locality and the associated
results on the connection between the spin and statistics and CPT invariance of the theory do not in general hold. Therefore,
the use of infinite-component fields to describe particles of finite spin, which arise naturally on the quantization of a harmonic
particle, presupposes calculation of the commutation function and an investigation of the locality of the theory on the basis of
it. It can be shown that locality can be achieved only for the correct connection between the spin and statistics at positive integer
points (when the integer or half-integer conformal weight $c\geq |s|$ -- the case of finite-component fields). Essentially complex
and noninteger or nonhalf-integer values of the conformal weight are excluded by the locality condition.

The commutation function
$$
  i\Delta_\chi(x-y;\zeta,\omega)=
  [\varphi_\chi(\zeta,x)\,,\,\varphi_\chi^+(\omega,y)]_\sigma ,
$$
can be calculated by means of the standard commutation relations for the particle creation and annihilation operators:
$$
[a(\mathbf{p},\lambda)\,,\,a^+(\mathbf{p}^\prime,\lambda^\prime)]_\sigma =
2p^0\delta(\mathbf{p}-\mathbf{p}^\prime)\delta_{\lambda \lambda^\prime}\,,\qquad
p^0>0\,,
$$
where the index $\sigma=\pm$ here and below denotes the (anti)commutator. We set
$$
\Delta_\chi(x;\zeta,\omega)=
\left(|\xi_\chi|^2\Delta_\chi^{(+)}(x;\zeta,\omega)+
|\tilde\xi_\chi|^2\Delta_\chi^{(-)}(x;\omega,\zeta)\right)/(2k)^{2{\rm Re}c}\,.
$$
In the positive-frequency part
\begin{equation}
i\Delta_\chi^{(+)}(x;\zeta,\omega)=
(2\pi)^{-3}\int d^4 p\delta_+(p^2) e^{ipx}\Pi_\chi(p;\zeta,\omega)
\end{equation}
the polarization function is
$$
\Pi_\chi(p;\zeta,\omega)=(\zeta\hat p\bar\zeta)^c(\omega\hat p\bar\omega)^{\bar c}
(\zeta\hat p\bar\omega)^s\overline{(\zeta\hat p\bar\omega)}^{-s}/|\Gamma(c+|s|+1)|^2\,.
$$
The negative-frequency part has the form
$$
\Delta_\chi^{(-)}(x;\zeta,\omega)=\sigma\Delta_{\bar\chi}^{(+)}(-x;\omega,\zeta)\,.
$$
The commutation function and its parts with purely positive or negative frequency are separately homogeneous of bidegree $\chi$
with respect to $\zeta$ and bidegree $\bar\chi$ with respect to $\omega$,
are Lorentz invariant, and satisfy the same equations (53) as the fields that
are commuted. The integrand in (54) is nonzero only at points of the future light cone: $p^2=0$, $p^0>0$, which is what gives
expression to the spectral condition. Therefore $\Delta_\chi^{(\pm)}(x;\zeta,\omega)$ as functions of $x$ are holomorphic in the past (future) tube
$$
T^{(\mp)}=\{ x\in \mathbb{C}^4| Im \,
x\in\stackrel{o}{V}_{(\mp)} \}\,,
$$
where $\stackrel{o}{V}_{(\mp)}$ is the interior of the future (past) light cone.
Therefore, it is sufficient to find an expression for $\Delta_\chi^{(+)}(x;\zeta,\omega)$ for
$x\in T^{(-)}$ with real timelike part and small imaginary ${\rm Im}x^0=-\epsilon$ and for the remaining points of the tube to use analytic
continuation. The commutation function for real points of ordinary space is the boundary value of the commutation function
in the past tube.

To calculate the integral (54), it is convenient to use a parametrization of the positive isotropic 4-momentum of the form
$$
p^\mu=p_\bot({\rm ch}a,{\rm cos}\varphi,{\rm sin}\varphi,{\rm sh}a)
$$
where $0<p_\bot<\infty$, $0\leq \varphi<2\pi$, and $a$ is real.
The calculation is readily done in a frame of reference in which $\zeta=(\alpha,0)$ and
$\omega=(0,\alpha)$. Transition to such a system is possible when $\zeta\omega\neq 0$.
The result admits the value $\zeta\omega= 0$, and therefore a special study
for this case is not needed.

The triple integral (54) reduces to the integral of a product of a Bessel function of integer order and a Macdonald function,
and the value of it when ${\rm Re}a>-1/2$ can be expressed in terms of a hypergeometric series. For ${\rm Re}a\leq -1/2$, one considers
a regularized value of the integral obtained by analytic continuation with respect to $c$ for fixed $s$. The upshot is that for $x$ in
the past tube with positive timelike part we have
\begin{equation}
i\Delta_\chi^{(+)}(x;\zeta,\omega)=
(\zeta\hat x\bar\zeta)^c(\omega\hat x\bar\omega)^{\bar c}
(\zeta\hat x\bar\omega)^s\overline{(\zeta\hat x\bar\omega)}^{-s}\,h_\chi(x^2,\nu) \,,
\end{equation}
where
\begin{equation}
h_\chi(x^2,\nu)=
(2\pi)^{-2}
e^{-3i\pi {\rm Re}c}
(x^2)^{-2{\rm Re} c-1)}(1+\nu^2 x^2)^{|s|}\,\,\times
\end{equation}
$$
 \qquad\qquad\qquad\qquad  {}_2 F_1(|s|-c,|s|-\bar c;2|s|+1;1+\nu^2 x^2)/\Gamma(2|s|+1)
$$
and
$$
\nu^2\equiv|\zeta\omega|^2/(\zeta\hat x\bar\zeta)(\omega\hat x\bar\omega)\,.
$$

To find the total commutation function at real space-time points
\begin{equation}
\Delta_\chi(x;\zeta,\omega)=
\lim_{x^\prime\rightarrow x}
\left\{|\xi_\chi|^2\Delta_\chi^{(+)}(x^\prime;\zeta,\omega)+
\sigma|\tilde\xi_\chi|^2\Delta_\chi^{(+)}(-\overline{x^\prime};\omega,\zeta)
\right\},
\end{equation}
where $x^\prime\in T^{(-)}$,
we make an analytic continuation of the function (55) through a spacelike region of the real part of $x$ to points $(-\bar x)$, keeping the
imaginary part of $x^0$ small. This route of the analytic continuation makes it possible to avoid, when this can be done, the
necessary cuts in the planes of the composite variables on which the multiply valued factors depend in the expression for
$\Delta_\chi^{(+)}(x;\zeta,\omega)$ written in the form
\begin{equation}
  i\Delta_\chi^{(+)}(x;\zeta,\omega)=
  |\xi_\chi|^2 \frac 1{(2\pi)^{2(1+Re c)}}
  e^{-i\pi(Re c+1)}|\zeta\omega|^{2 Re c}e^{2is\arg(\zeta\hat x\bar\omega)}
\,\,\times
\end{equation}
$$
\qquad\qquad\qquad\qquad\qquad
\left(\frac{\zeta\hat x\bar\zeta}{\omega\hat x\bar\omega}\right)^{i Im c}
  (\nu^{-2})^{-(Re c+1)}(1+\nu^2 x^2)^{|s|} \,\,\times
$$
$$
\qquad\qquad\qquad {}_2 F_1(c+|s|+1,\bar c+|s|+1;2|s|+1;1+\nu^2 x^2)/\Gamma(2|s|+1)\,.
$$
Here the last four factors, which depend on $(\zeta\hat x\bar\zeta)/(\omega\hat x\bar\omega)$, $\nu^{-2}$, and $1+\nu^2 x^2$, are multiply valued.
In the frame of reference mentioned above,
$$
(\zeta\hat x\bar\zeta)/(\omega\hat x\bar\omega)=
\nu^2(x^0+x^3)^2$$
$$
\nu^{-2}=x_\bot^2-x^2\,, \qquad 1+\nu^2 x^2=\nu^2 x_\bot^2
$$
where $x_\bot^2=x_1^2+x_2^2$.

We take the cuts in the planes of these variables, when they are needed, along the positive real half-axes for the first three
factors, and for the hypergeometric function from $+1$ along the positive real half-axis to $\infty$. The route of the analytic
continuation remains on one side of the cut in the plane of the variable $(\zeta\hat x\bar\zeta)/(\omega\hat x\bar\omega)$ and passes round the cuts in the planes of the remaining variables.

When the branch points of the solution of the hypergeometric equation are avoided, it becomes a linear combination of
two independent solutions. Only in the degenerate case when among the parameters of the hypergeometric function
$F(\alpha,\beta;\gamma; z)$ one of the numbers $\alpha$, $\beta$, $\gamma-\alpha$, $\gamma-\beta$
is an integer does the solution go over into itself, apart from a constant factor.
Therefore, vanishing of the commutation function (57) for spacelike $x$ is possible if we consider representations with positive
real part of the conformal weight $c$ only when the number $|s|-c$ is an integer. As examination of all possible cases of
degeneracy shows [39], degeneracy of the hypergeometric series in (58) corresponds to nonpositive values of the indicated
number, i.e., to integer positive homogeneity index $\chi$.

Beating in mind now that
$${\rm arg}(\omega(-\hat{\bar x})\bar\zeta)={\rm arg}(\zeta\hat{x}\bar\omega)\pm\pi\,,$$
we find that the total commutation function (57) in the spacelike region is proportional to
$$
|\xi_\chi|^2+\sigma e^{2\pi is}|\tilde\xi_\chi|^2\,.
$$
Under the restriction that we have made on $c$, the product
$$
(\nu^{-2})^{-(Re c+1)}(1+\nu^2 x^2)^{|s|}
$$
is single valued in this region. Therefore, the requirement of locality (microcausality) leads to the ``crossing'' condition [18]
$$
|\xi_\chi|=
|\tilde\xi_\chi|
$$
and the usual connection between the spin and statistics:
$$
(-1)^{2s}=-\sigma\,.
$$
Redefining the phases of the operators $a(\mathbf{p})$ and $b(\mathbf{p})$ and changing the normalization of the field $\varphi_\chi(\zeta,x)$,
we can assume without loss of generality that
$$
\xi_\chi=\tilde\xi_\chi=1\,.
$$

Thus, the obtained set of massless nongauge quantum fields for which locality holds and, as a consequence, there is the
usual connection between the spin and statistics includes only finite-component fields corresponding to integer positive
homogeneity index $\chi$.

\vspace{1cm}

We are grateful to I.A.\,Bandos for numerous helpful discussions, and also to the participants of the seminar of D.V.\,Volkov.

\end{document}